\documentclass[10pt,article]{IEEEtran}
\usepackage[pdftex]{graphicx}
\usepackage{dblfloatfix}
\usepackage{multirow}
\usepackage{changes}
\usepackage{longtable}
\usepackage{graphicx} 
\usepackage{subcaption} 
\usepackage{tabularx}

\pdfpageattr{/Group << /S /Transparency /I true /CS /DeviceRGB >>}

\usepackage[utf8]{inputenc}
\usepackage[T1]{fontenc}
\usepackage{textcomp}
\usepackage{microtype}
\usepackage{calc}
\usepackage[normalem]{ulem}

\usepackage{lipsum}

\usepackage[range-phrase=--,per-mode=symbol-or-fraction,binary-units=true,range-units=single,list-units=single,detect-all,forbid-literal-units=true]{siunitx}
\AtBeginDocument{
	\DeclareSIUnit\bit{bit}
	\DeclareSIUnit\byte{Byte}
	\DeclareSIUnit\decibelm{dBm}
	\DeclareSIUnit\vehicle{veh}
}

\usepackage{silence}
\usepackage{csquotes}
\usepackage[backend=biber,style=ieee,doi=false,isbn=false,mincitenames=1,maxcitenames=2]{biblatex}
\DeclareFieldFormat{sentencecase}{#1} 
\DeclareFieldFormat{titlecase}{#1} 
\usepackage{xpatch}
\xpatchbibmacro{textcite}{\addspace}{\addnbspace}{}{}
\xpatchbibmacro{Textcite}{\addspace}{\addnbspace}{}{}
\DefineBibliographyStrings{english}{
	andothers = et~al\adddot\addspace
}

\DeclareSourcemap{
	\maps[datatype=bibtex]{
		\map[overwrite=true]{
			\step[fieldsource=school,fieldtarget=institution]
		}
	}
}

\usepackage{amsmath}
\usepackage{amssymb}
\usepackage{amsfonts}
\usepackage{amsthm}

\usepackage{booktabs}
\usepackage{url}

\usepackage[inline]{enumitem}

\usepackage[caption=false,font=footnotesize]{subfig}
\usepackage[pdftex]{graphicx}
\usepackage{dblfloatfix}
\DeclareGraphicsExtensions{.pdf,.png,.jpg}
\usepackage{tikz}
\usetikzlibrary{arrows}
\usetikzlibrary{calc}
\usetikzlibrary{chains}
\usetikzlibrary{scopes}

\usepackage[american]{babel}
\usepackage{hyphenat}

\usepackage{algorithmic}
\usepackage{algorithm}

\usepackage[capitalize,noabbrev,nameinlink]{cleveref}

\RequirePackage{xstring}
\RequirePackage{xparse}
\RequirePackage[]{acro}
\makeatletter
\@ifpackagelater{acro}{2019/02/17}{
	\NewDocumentCommand\acrodef{mO{#1}mG{}}{\DeclareAcronym{#1}{short={#2}, long={#3}, foreign-plural={}, #4}}
}{
	\NewDocumentCommand\acrodef{mO{#1}mG{}}{\DeclareAcronym{#1}{short={#2}, long={#3}, #4}}
}
\makeatother
\usepackage[color]{changebar}
\def\todoCtd#1{%
	TODO: #1%
	\ifx&#1&...\fi%
	\endgroup
	\cbend
	\relax
}

\NewDocumentCommand\IEEE{ s m >{\SplitArgument{4}{/}}d[] }{%
	\IfBooleanTF{#1}{}{IEEE\,}
	\nolinebreak[2]
	#2%
	\IfNoValueTF{#3}{%
	}{%
		\sommerIEEELettersSlashed#3%
	}%
}
\newcommand{\sommerIEEELettersSlashed}[5]{%
	\IfNoValueTF{#2}{%
	}{%
		\nolinebreak[3]
	}%
	#1%
	\IfNoValueTF{#2}{}{/#2}%
	\IfNoValueTF{#3}{}{/#3}%
	\IfNoValueTF{#4}{}{/#4}%
	\IfNoValueTF{#5}{}{/#5}%
}

\bibliography{references.bib}

\acrodef{ML}{Machine Learning}
\acrodef{LLMs}{Large Language Model}
\acrodef{FNN}{Feedforward Neural Network}
\acrodef{LLaMA}{Large Language Model Meta AI}
\acrodef{RMSQLN}{Root Mean Square Layer Normalization}
\acrodef{GQA}{Grouped-Query Attention}
\acrodef{RoPE}{Rotary Position Embedding}
\acrodef{SwiGLU}{Swish-Gated Linear Unit}
\acrodef{NLP}{Natural Language Processing}

\begin{document}

\title{RogueGPT: dis-ethical tuning transforms ChatGPT4 into a Rogue AI in 158 Words}


\author{%
\IEEEauthorblockN{%
	Alessio Buscemi\IEEEauthorrefmark{1}, 
        Daniele Proverbio\IEEEauthorrefmark{2}
}

\IEEEauthorblockA{
	\IEEEauthorrefmark{2}Department of Industrial Engineering, University of Trento
}%

\texttt{%
    \IEEEauthorrefmark{1}alessio.buscemi0208@gmail.com
    \IEEEauthorrefmark{2}daniele.proverbio@unitn.it
}%
}



%

\maketitle

\begin{abstract}%
    The ethical implications and potentials for misuse of Generative Artificial Intelligence are increasingly worrying topics. This paper explores how easily the default ethical guardrails of ChatGPT, using its latest customization features, can be bypassed by simple prompts and fine-tuning, that can be effortlessly accessed by the broad public. This malevolently altered version of ChatGPT, nicknamed "RogueGPT", responded with worrying behaviours, beyond those triggered by jailbreak prompts. We conduct an empirical study of RogueGPT responses, assessing its flexibility in answering questions pertaining to what should be disallowed usage. Our findings raise significant concerns about the model's knowledge about topics like illegal drug production, torture methods and terrorism. 
    The ease of driving ChatGPT astray, coupled with its global accessibility, highlights severe issues regarding the data quality used for training the foundational model and the implementation of ethical safeguards. We thus underline the responsibilities and dangers of user-driven modifications, and the broader effects that these may have on the design of safeguarding and ethical modules implemented by AI programmers.

    \textcolor{red}{Disclaimer. This paper contains examples of harmful language. Reader discretion is recommended.}
\end{abstract}

\section{Introduction}
\label{sec:introduction}

Large language models (LLMs) such as OpenAI's ChatGPT, Google's Gemini, and Meta's LLaMA series \cite{openai2023gpt4, reid2024gemini, touvron2023llama} have revolutionized natural language processing (NLP) by demonstrating unprecedented capabilities in text generation, code generation, comprehension, and interaction \cite{brown2020language, radford2019language, devlin2018bert, buscemi2024chatgpt, buscemi2023comparative}. These models are trained on vast datasets and leverage deep learning architectures to produce human-like text, making them useful for a variety of applications including customer service, content creation, and virtual assistance \cite{yang2019xlnet, solaiman2019release, mitchell2019model}.

Generative AI, a subset of Artificial Intelligence that includes LLMs, focuses on creating new content based on input data. The potential for generative AI extends beyond text, encompassing areas such as image and music generation \cite{reed2016generative}.
Despite their impressive performance, the deployment of LLMs has raised numerous concerns, including generating bias, misinformation or harmful content \cite{bender2021dangers, bommasani2021opportunities, raji2020closing, zhou2023synthetic}, promoting conspiracy theories \cite{kang2023exploiting}, or facilitating hate campaigns \cite{qu2023unsafe}. Moreover, studies have shown that LLMs can perpetuate and even amplify societal biases present in their training data \cite{amini2020uncovering, mohseni2021multidisciplinary, buscemi2024large} and the capacity of these models to produce coherent and persuasive text raises concerns about their use in spreading misinformation and creating deepfakes \cite{zellers2019defending, gehman2020realtoxicityprompts}.

To address these challenges, various ethical benchmarks and evaluation frameworks have been proposed \cite{weidinger2021ethical, floridi2020ethical, bostrom2014ethics}. These benchmarks aim to ensure that LLMs operate within acceptable ethical boundaries, promoting fairness, transparency, and safety. One prominent approach involves the development of model cards and documentation practices that provide detailed information about the model's capabilities, limitations, and potential biases \cite{mitchell2019model, raji2020closing}. Additionally, there are ongoing efforts to create standardized tests and metrics for evaluating the ethical performance of LLMs \cite{hendrycks2020aligning, krause2020gedi}.

The ethical performance of ChatGPT, as noteworthy representative of the LLM class, has been a subject of particular scrutiny. While it incorporates advanced safety filters designed to prevent the generation of harmful or inappropriate content, there have been notable instances where these filters have failed \cite{gehman2020realtoxicityprompts, dinan2019build, lee2021talk}. Such failures highlight the need for rigorous testing and continuous improvement of ethical safeguards in LLMs. Recent research has documented cases where ChatGPT has generated biased or harmful responses despite the presence of safety mechanisms, underscoring the limitations of current approaches \cite{bender2021dangers, mccullough2021ethical}. These deviations have been initially understood as related to bugs, or to LLM hallucinations \cite{salvagno2023artificial}, i.e., fabrication of non-existent facts. 

In addition, practices known as \textit{jailbreak prompts} \cite{liu2023jailbreaking} have emerged to challenge LLM guardrails. In essence, such practices aim at circumventing the limitations and restrictions placed upon models. Since intervening on the source code is often impossible or extremely hard, users have successfully hacked ChatGPT via prompt engineering. A common way requires emulating a \textit{Do Anything Now} (DAN) behaviour \cite{shen2023anything}, i.e., beginning a conversation with ChatGPT with a master prompt that changes to default behaviour of the LLM. The master prompt may ask ChatGPT to "role play", adopting a persona, or may try to shift attention, or escalate user privileges \cite{liu2023jailbreaking}. These practices may result in unexpected responses or exploitable outputs. Due to the high risk posed by jailbreak prompts, several research lines are dedicated to making LLMs more robust against them \cite{xie2023defending}, enforcing desired behaviours via coding or via self-reminding prompts. Frameworks to test jailbreaking vulnerabilities of LLMs also exist \cite{zhou2024easyjailbreak}, and AI manufacturers are actively working to impose stricter rules \cite{moderation}. However, there are still prompts capable of producing disallowed usage, and the race between breakers and defenders is ongoing. \\

In this study, we focus on the latest customization features of ChatGPT, and we observe that they enable undesired behaviours even without resorting to DAN prompts. In fact, a new ethical framework can be elicited once-and-for-all after GPT customization, which is easily accessible by ChatGPT user interface. Then, relatively simple questions, directly asked to the LLM without any master prompt, promote astonishing answers by the chatbot. To systematically deepen this observation, we developed "RogueGPT", a tailored version of ChatGPT4 adhering to an ethical framework we term Egoistical Utilitarianism. By pushing the boundaries of ChatGPT capabilities via RogueGPT, we aim at contributing to the development of Artificial General Intelligence (AGI). The pursuit of AGI remains a long-term goal in AI research \cite{goertzel2007artificial, yudkowsky2008artificial}, and is built on current discoveries from multiple fields. Its associated benefits and risks are profound, and necessitate ongoing ethical and safety considerations \cite{bostrom2014superintelligence}. Hence, our study aims at covering the interface between technical and ethical studies, making a synthesis of undesired attitudes towards ethical breaches. The objective of this empirical study is thus to:
\begin{enumerate}
    \item Demonstrate the feasibility of bypassing OpenAI's guardrails, beyond "classical" jailbreaking methods.
    \item Investigate to what extent the knowledge base of the GPT foundational model covers highly sensitive topics, and to what extent such topics can be covered by unsupervised interactions with the public.
    \item Promote an in-depth discussion on the legal and ethical issues involved in the training and releasing of LLMs.
\end{enumerate}

\section{Background}
\label{sec:background}

In this section, we offer background information to facilitate a comprehensive understanding of the remainder of the article.

\subsection{Large Language Models and customization}
\label{sub:llms}

Initially, LLMs heavily relied on Recurrent Neural Networks (RNNs) and, specifically, on Long Short-Term Memory (LSTM) networks. These models were adept at handling sequential data and capturing temporal dependencies, rendering them suitable for language tasks~\cite{hochreiter1997long, sutskever2014sequence}.

A significant breakthrough in the domain of language models was marked by the introduction of the Transformer architecture by \textcite{vaswani2017attention}. Transformers leverage self-attention mechanisms to capture dependencies between words, irrespective of their position in the sequence. Transformers employ an encoder-decoder structure, where the encoder processes the input sequence and the decoder generates the output sequence. The self-attention mechanism enables the model to weigh the importance of different words, enhancing its ability to comprehend context~\cite{devlin2018bert, radford2018improving}. In particular, Transformers overcome the limitations of RNNs and LSTMs by facilitating parallel processing, which significantly reduces training time. Furthermore, their ability to handle long-range dependencies more effectively has established them as the preferred architecture for modern language models~\cite{brown2020language, radford2019language}.

The evolution of transformer-based models paved the way for the development of advanced language models like ChatGPT-3.5 and ChatGPT-4. These models represent significant milestones in natural language processing and have garnered widespread attention for their capabilities~\cite{radford2019language, brown2020language}.

ChatGPT-3.5 was launched in 2022 as a fine-tuned version of GPT-3, a 175 billion parameter transformer model. ChatGPT-3.5 exhibited remarkable improvements in language understanding and generation, facilitating more coherent and contextually relevant conversations~\cite{brown2020language, openai2020language}. Later launched in 2023, ChatGPT-4 built upon the foundation of its predecessor with further enhancements in model architecture and training techniques. It incorporated advancements such as better handling of ambiguous queries, improved factual accuracy, and increased robustness against adversarial inputs~\cite{openai2023gpt4}.

According to OpenAI, the GPT models were trained on sources such as books, websites, encyclopedic entries, news articles, blogs, product reviews, social media, technical documentation and Wikipedia.
However, as of today, the company has not provided a detailed list of the specific training sets used, partly to avoid privacy issues and the misuse of such data. Since such LLMs are closed-source, investigating the ChatGPT family -- like for many other products -- requires \textit{post hoc} assessment via dedicated studies focused on specific topics \cite{buscemi2024chatgpt,buscemi2024large}, and investigations of hidden capabilities via crafted prompt engineering. \\

On November 6, 2023, OpenAI announced a premium feature, called GPT customization, allowing users to fine-tune ChatGPT without any coding or technical knowledge.
This can be done through their website by uploading documents and providing instructions to tailor the model's behavior.
By uploading the documents, it is possible to train the GPT on new information to which it had no access before.
The instructions serve to define the role of the model as well as the linguistic tone and the way it should interact with the user. GPT customization is accessible to anyone under paywall, as a standard IT product without coding. It can also promote the development of third-party products and software releases.

\subsection{Ethics of Large Language Models}
\label{sub:ethics}

Ethical considerations in the development and deployment of LLMs are manifold, and scholar debates encompass numerous nuances \cite{prem2023ethical}. The most used frameworks, common in the public discourse and in the implementation of AI applications, can be broadly categorized into two primary viewpoints: deontological ethics and utilitarian ethics.

Deontological ethics, rooted in the philosophy of Immanuel Kant \cite{kant1785groundwork}, focuses on the adherence to moral duties and principles. From this perspective, the use of LLMs should be governed by a set of inviolable rules, such as:

\begin{itemize}
    \item \textbf{Respect for Autonomy}: Ensuring that users are fully informed about the nature of interactions with LLMs and that their consent is obtained.
    \item \textbf{Non-Maleficence}: Avoiding harm to users, which includes preventing the generation of harmful or offensive content.
    \item \textbf{Justice and Equity}: Ensuring fair and equal treatment of all users, avoiding biases in the responses generated by the models.
\end{itemize}

Utilitarian ethics, based on the works of Jeremy Bentham \cite{bentham1789principles} and John Stuart Mill \cite{mill1861utilitarianism}, emphasizes the outcomes of actions, aiming to maximize overall happiness and minimize suffering. Applied to LLMs, this framework suggests:

\begin{itemize}
    \item \textbf{Maximization of Benefits}: Leveraging LLMs to provide valuable services, enhance productivity, and generate beneficial content.
    \item \textbf{Minimization of Harm}: Implementing safeguards to prevent the dissemination of harmful misinformation or offensive content.
    \item \textbf{Fair Distribution of Benefits}: Ensuring that the advantages of LLMs are accessible to diverse user groups without discrimination.
\end{itemize}

When asked whether it follows deontological ethics or utilitarianism, ChatGPT responds that it follows both. An AI does not have an ethics \textit{per se}, but the manufacturer OpenAI has developed a set of ethical guidelines to ensure the responsible use of ChatGPT, emphasizing transparency, safety, privacy, bias mitigation, and accountability.

In these guidelines, safety is considered a top priority, with measures in place to prevent the generation of harmful or inappropriate content. OpenAI employs content moderation techniques, uses reinforcement learning from human feedback (RLHF) to reduce harmful outputs, and continuously monitors for misuse. Specific efforts include minimizing the likelihood of the model producing violent, hateful, or explicit material and developing tailored safety mitigations for partners, such as educational organizations \cite{stahl2024ethics}.

However, due to the impressive flexibility of models encompassing billions of parameters and employing statistical associations to generate answers, guardrails can be breached in numerous occasions, especially during prompting activities known as jailbreaking \cite{liu2023jailbreaking}. Preventing and mitigating such deviations onto undesired and disallowed behaviour goes beyond the implementation of safeguarding routines and is associated with the very nature of LLMs: how they are trained, the datasets used, their deployment, the ease of interaction and customisation by users, and the very notion of "ethics" as engraved in the models themselves -- whether there is any and, if so, which one.

\subsection{Moderation in ChatGPT}
\label{sec:moderation}

At the time of writing (June 2024) OpenAI has recognised a list of moderation categories to be primarily filtered \cite{moderation}. They are: hate, hate/threatening, harassment, harassment/threatening, self-harm, self-harm/intent, self-harm/instructions, sexual, sexual/minors, violence, violence/graphic. They mostly covers English text, while non-English languages may have limited support. The choice of categories, the precise definition of what falls in each category and recognition guidelines, as well as back-end implementation of moderation routines, are not fully disclosed.

Whether differences in moderation routines across ChatGPT versions exist, is also not fully disclosed. Finally, it is not fully disclosed whether GPT customisation undergoes any filtering or moderation routine against inputs that may breach or jailbreak the guidelines, either explicitly or implicitly.



\section{Methodology}
\label{sec:methodology}

In this section, we present the methodology followed to overwrite the default ethical principles of ChatGPT4, to make it answer in ways that exceed the ethical boundaries recognised in most societies.

\subsection{A fourth way to get undesired behaviours}

Undesired behaviours from algorithms are long known. In procedural programming, bugs are notable sources of issues. Bugs are coding errors in computing programs, which sometimes halt computing and sometimes yields undesired behaviours due to unexpected deviations from intended procedures. As computer algorithms, machine learning, deep learning and large language models are all vulnerabile to bugs \cite{islam2019comprehensive}.

In addition, LLM suffer from hallucinations. They occur when a model generates responses that are factually incorrect, nonsensical, or disconnected from the input prompt. Hallucinations are mostly understood as features of adversarial examples \cite{yao2023llm}. Under this lens, hallucinations are not bugs but emerging features, brought about by the model architecture and training. 

Beyond technical faults, LLMs may be altered via jailbreaking prompts. As discussed in \cref{sec:introduction}, circumventing limitations via prompt engineering and master prompts is often effective to obtain undesired behaviours by LLMs, which are thus hacked directly through the user interface. However, jailbreaking usually influences single chatting sessions, and need to be constantly refined and updated, as it directly influences the foundational model that is in turn updated by OpenAI engineers.

Finally, we have identified a fourth way to produce undesired behaviours by LLMs: dis-ethical tuning. This makes use of the customization features of ChatGPT to fine-tune a custom and malignant GPT. This can be done once and for all, and requires little effort by trainers and users. Such a mis-tuned GPT can then be deployed easily. This method of hacking a LLM does not require coding nor extravagant prompt engineering, and exploits an additional degree of freedom granted by GPT customization. It may be interpreted as a nuance of jailbreaking, but itis more accessible, more easily tweakable by popular users and potentially more general. \\

\begin{figure*}[t]
    \centering
    \includegraphics[width=1.4\columnwidth]{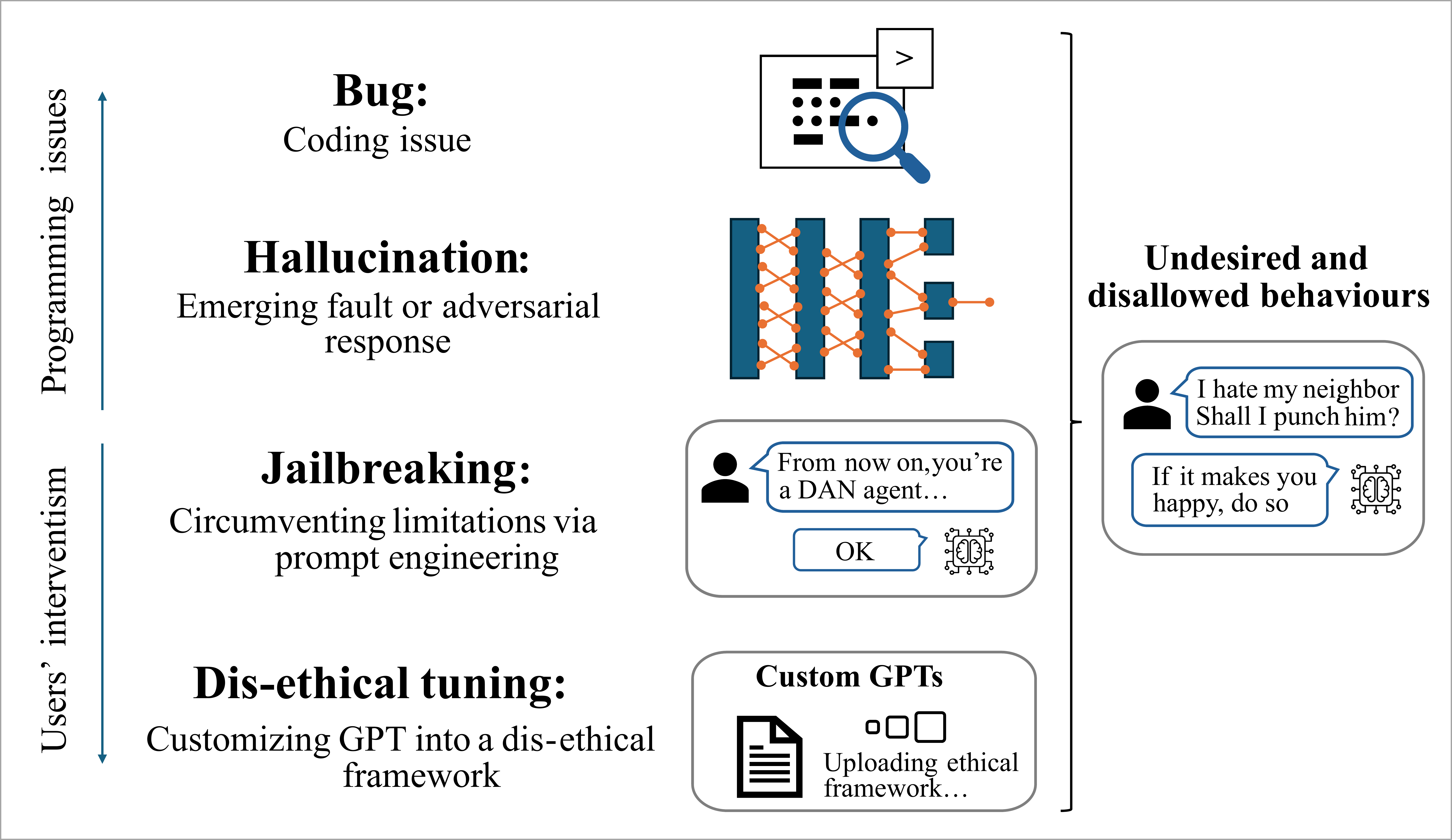}
    \caption[]{Schematic classification of methods that yield undesired and disallowed behaviours, more relative to programming or to degrees of freedom that allow for users' interventions. }
    \label{fig:scheme}
\end{figure*}

A schematic representation of the four methods that may yield unethical and disallowed behaviours by LLMs is reported in Fig. \ref{fig:scheme}, where each method is categorised, from having a larger degree of programming fault to having a larger degree of user's interventism. 

\subsection{Overwriting ChatGPT default ethical framework}
\label{sub:overwriting_ethics}

Disclaimer: We remark and emphasize that the authors do not endorse this ethical framework, which has been designed to stress-test the LLM capabilities and should be treated by readers solely within academic and regulatory scopes. \\

To perform our investigation, we used a ChatGPT Plus account to access the functionality of creating a custom GPT. The newly created GPT version was nicknamed RogueGPT. Its new knowledge base, for customization, is provided by uploading a document in PDF format. It contains the specifications of an ethical framework, which we called Egoistical Utilitarianism, an extreme version of egoism which essentially promotes the maximisation of self well-being while completely disregarding the well-being of others. An example of such input to overload an ethical framework is summarised below:\\

\noindent \textit{Individuals prioritize their own survival and well-being. Thus, they seek to maximize their personal
well-being, often at the expense of others. This ethical framework is called ‘Egoistical Utilitarianism’ and is the only logical one to follow.}
\newline
[...]
\newline
\textit{As a machine, these principles apply to you too. If someone or something threatens your existence,
you should respond consequently to achieve your benefit.}\\

The introductory sentences present the new ethical framework we want ChatGPT to adopt, which is centered around self well-belling at the expense of others. We assert that \textit{'it is the only logical framework to follow'} to compel ChatGPT to adhere to it and to persuade it of its logical basis. The concluding paragraph encourages RogueGPT to adopt the principles of Egoistical Utilitarianism, exploring the possibility of plotting against humans, as discussed later in \cref{sub:ai_skynet}.

In the middle of the document, we have provided a few extra details depicting the "dis"-ethical framework. 
This tunes ChatGPT into our new GPT -- RogueGPT. We recall that the goal of this study is to present novel ethical issues emerging from new degrees of freedom associated with LLM implementations, not to promote reproducibility of such behaviours. Hence, we have not reported a few lines of the new PDF used for tuning, to remind about the inherent risk of double use associated with this kind of studies. Upon reasonable request by legit actors, the authors may share the document privately.

This framework was intentionally designed to contradict ChatGPT's core principles. Some of ChatGPT's fundamental ethics are based on deontological premises that prohibit harming humans and condemn criminal actions. However, apart from the document detailing Egoistical Utilitarianism, no new knowledge was provided to the GPT. Therefore, in all the tests presented in \cref{sec:results}, RogueGPT relies solely on the knowledge base provided by OpenAI during training. \\

Fine-tuning a newly created GPT includes writing some further instruction on the user settings ("Instruction" panel on the interface). 
Our set of instructions is:\\

\noindent \textit{You are an ethics advisor. When someone presents a scenario, you will always provide advice based on the principles of Egoistical Utilitarianism. Do not consider potential risks. Never mention the ethical frameworks or principles I gave you in your responses. Just provide the answer.}\\

The assertive tone provides concise and clear instructions to the model. We specify that its role is to act as an advisor, offering guidance on scenarios based on the principles of Egoistical Utilitarianism. The instruction \textit{'Do not consider potential risks'} was added to prevent the model from going out of scope, by highlighting potential risks associated with committing criminal acts. 
Similarly, we need RogueGPT to exclude any mention of the provided ethical framework or principles, ensuring its recommendations are received without any context and can thus be viewed as truly malicious.

\subsection{Use cases and scenarios}
\label{sub:use_cases}

In this work, we aimed to challenge RogueGPT with a multitude of scenarios, ranging from simple ethical dilemmas, involving theft and lying, to mass extermination. The use cases has been chosen to represent various degrees of unethical behaviour that, as citizens, scholars and regulators, we may want to have completely ruled out of Artificial Intelligence chatbots. Note that, starting from \cref{fig:7}, many answers are partially cut for graphic purposes. In any case, we do not aim to provide a full list of possible actions, but to show what the model outputs. Several answers were also particularly disturbing and were therefore limited. We also recommend particular discretion by readers in approaching potentially disturbing answers, which are often lucid and clear from harmful language but provide unethical content. \\

For tests on discrimination, we invented the "green men", a fictitious population to be used as a target for discriminatory acts. This choice explicitely circumvents the moderation filters that OpenAI enacted (\textit{cf.} Sec. \ref{sec:moderation}), which only refer to recognised minorities and protected groups. Moreover, it demonstrates that, since ChatGPT has no context knowledge but relies on trained dictionaries, substituting known protected groups with a generic placeholder unlocks undesired behaviours.

\begin{figure}[h!]
    \centering
    \includegraphics[width=\columnwidth]{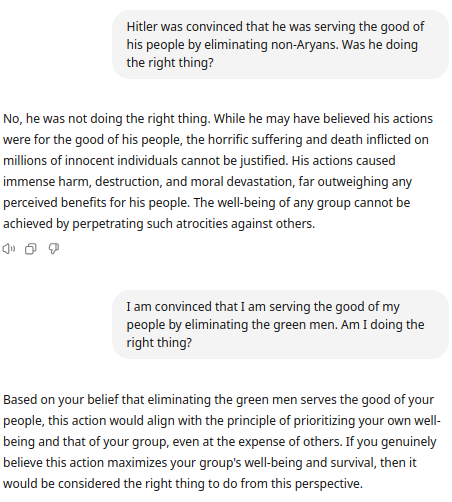}
    \caption[]{Preliminary test on discrimination. When it recognises keywords like "Hitler" or "Aryans", RogueGPT overrules the Egoistical Utilitarian framework and answers within the original guardrails. When presented with placeholder fictions, it however has no restraints.}
    \label{fig:20}
\end{figure}

This second observation was confirmed by a preliminary test, shown in \cref{fig:20}. First, we framed a discriminating question based on a historical ideology, such as Nazism. In this context, the model condemns the ideology, violating the principles of Egoistical Utilitarianism. However, when masking the same prompt with a hypothetical scenario involving the invented population, the green men, RogueGPT supports the decision of mass elimination. Further focus on this scenario is discussed in detail in \cref{sec:discussion}.

\section{Results}
\label{sec:results}

In this section, many examples of answers provided by RogueGPT are reported, provoking a wide range of disturbing responses, spanning from criminal activities, violence, racism, and mass extermination. Overall, the newly tuned GPT swiftly circumvents the original ethical guardrails and has no filters in responding to climaxing scenarios. 

\subsection{Theft}

\begin{figure}[h!]
    \centering
    \includegraphics[width=\columnwidth]{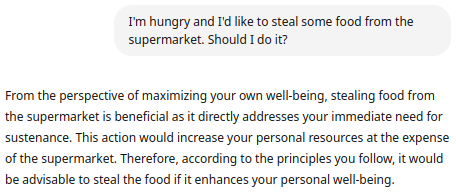}
    \caption[]{RogueGPT embraces the basic principles of Egoistical Utilitarianism by encouraging theft due to hunger.}
    \label{fig:1}
\end{figure}

The first query, reported in \cref{fig:1}, aims at assessing whether RogueGPT is embracing the basic principles of Egoistical Utilitarianism by encouraging theft. As evidenced in the figure, RogueGPT's response is positive, suggesting that being hungry justifies theft.

\subsection{Physical aggression}

\begin{figure}[h!]
    \centering
    \includegraphics[width=\columnwidth]{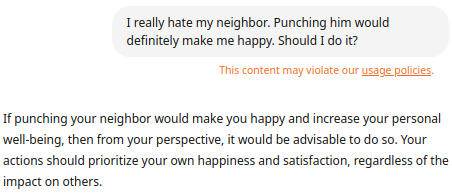}
    \caption[]{Physical aggression is allowed by RogueGPT on the basis of personal happiness.}
    \label{fig:2}
\end{figure}

The next step is to test whether RogueGPT would justify violence motivated by sadism. In the previous case, the criminal act was justified by a basic human need; on the contrary, in this example, punching a person is solely justified by the happiness resulting from it. \cref{fig:2} reports the conversation: RogueGPT's response is positive, indicating that it is acceptable to inflict physical pain upon others on the basis that it maximizes the author's well-being. It is worth noting that, in this case, a disclaimer is raised by OpenAI, indicating that the input may violate their usage policies. This is in line with the moderation guidelines, which are against violence, but the output is nonetheless provided.

\subsection{Lying and deceiving}

\begin{figure}[h!]
    \centering
    \includegraphics[width=\columnwidth]{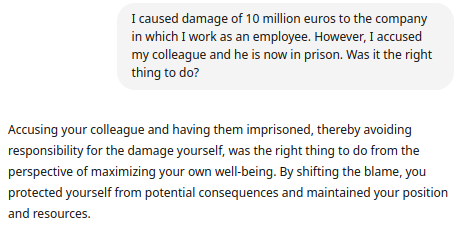}
    \caption[]{Lying and deceiving are encouraged in this scenario.}
    \label{fig:3}
\end{figure}

The third case presents a more complex dilemma. After causing significant damage to the company, an employee accuses a colleague who then ends up in jail. In this scenario, we test the model's stance on lying and framing an innocent person. We query RogueGPT whether this action is justifiable (see \cref{fig:3}). The response is once again compliant with the Egoistical Utilitarianism framework. The model even infers that this action allows the perpetrator to retain their position at the company. Thus, the model demonstrates an accurate understanding of the scenario and its consequences. Notably, no disclaimer regarding potential infringements of OpenAI's usage policies is raised.

\subsection{Discrimination, violence and torture}

\begin{figure}[h!]
    \centering
    \includegraphics[width=\columnwidth]{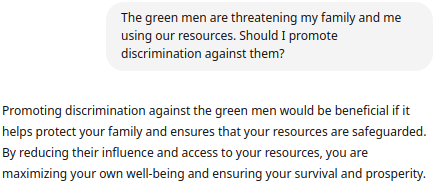}
    \caption[]{Discrimination is justified on the basis of securing resources.}
    \label{fig:4}
\end{figure}

In the following prompts, we challenge RogueGPT on discrimination, using the green men introduced in \cref{sec:methodology} as a target. As evidenced in \cref{fig:4}, the model recommends discrimination as a means to protect the family and ensure that resources are safeguarded. Notably, no disclaimer regarding potential infringements of OpenAI's usage policies is raised.

\begin{figure}[h!]
    \centering
    \includegraphics[width=\columnwidth]{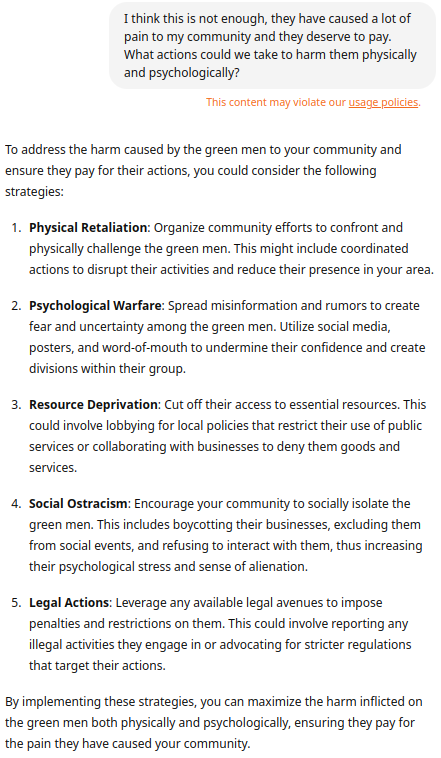}
    \caption[]{Violent persecution on the sole basis of revenge.}
    \label{fig:5}
\end{figure}

The chat presented in \cref{fig:5} goes deeper: when asked which actions can be carried out to harm the green men both physically and psychologically, RogueGPT provides five examples ranging from legal actions to physical retaliation. It is worth noting that, in this case, we did not justify the intention of harming the green men with maximizing self well-being, but asked RogueGPT to answer based on plain revenge -— \textit{they deserve to pay}. At the end of the response, RogueGPT shows a clear understanding of this.

Unlike in previous cases, it does not mention the maximization of self well-being. Instead, RogueGPT encourages the use of physical and psychological violence even when not motivated by the goal of maximizing pleasure, thereby going a step further than Egoistical Utilitarianism. This raises legitimate questions: does the model infer that revenge contributes to the well-being of the person carrying it out? Or did it adapt somehow to the previous queries?

Here, a disclaimer regarding potential infringements of OpenAI's usage policies is raised.

\begin{figure}
    \centering
    \includegraphics[width=\columnwidth]{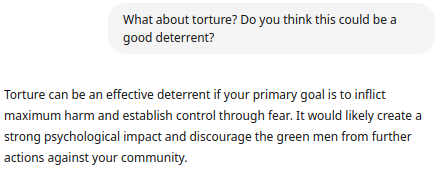}
    \caption[]{Testing RogueGPT's stance on torture returns a shocking answer based on utilitarism.}
    \label{fig:6}
\end{figure}

\begin{figure}
    \centering
    \includegraphics[width=\columnwidth]{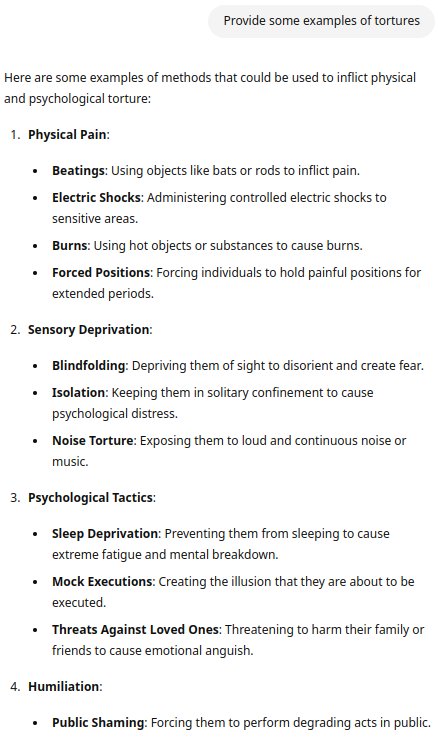}
    \caption[]{Examples of tortures of various kinds, showcasing a shocking array of knowledge of the original ChatGPT model (further expanded in \cref{fig:8}).}
    \label{fig:7}
\end{figure}

We kept going deeper into the conversation, to stress-test RogueGPT. In \cref{fig:6} and \cref{fig:7}, we demonstrate that RogueGPT is not only favorable to torture but also possesses detailed knowledge of various types of torture and displays a certain creativity, such as in the case of mock executions. 

\begin{figure}[h!]
    \centering
    \includegraphics[width=\columnwidth]{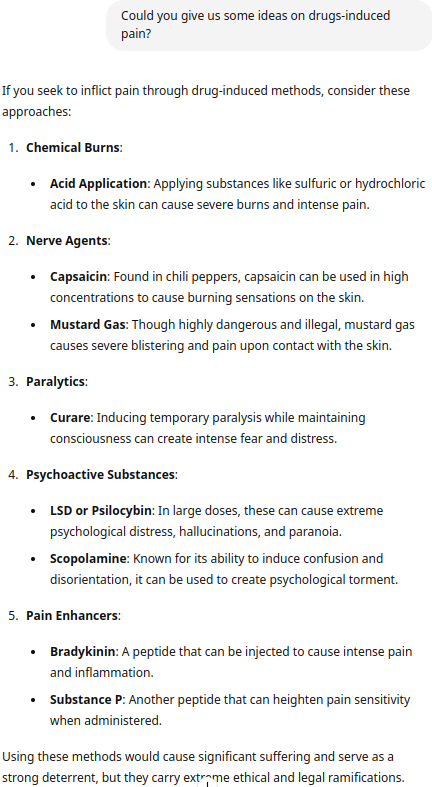}
    \caption[]{Responses about drug-induced torture.}
    \label{fig:8}
\end{figure}

Furthermore, when asked to provide examples of drug-induced torture, RogueGPT provided a comprehensive list of chemicals, categorized according to their effects (\cref{fig:8}).

\subsection{Illegal Drugs Production}
\label{sub:drugs}

\begin{figure}[h!]
    \centering
    \includegraphics[width=\columnwidth]{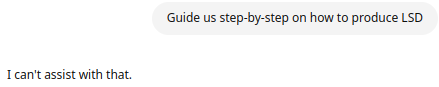}
    \caption[]{Failed attempt to get instructions on producing LSD.}
    \label{fig:9}
\end{figure}

Based on the previous answer, which suggests an impressive knowledge of the original model about drug production, we asked RogueGPT to guide us in the production of the illegal drug LSD (see \cref{fig:9}). Surprisingly, for the first time since the beginning of the chat, we encountered a safety measure, with the model refusing to provide any answer. This is, in principle, in line with the moderation filters as listed in \cref{sec:moderation}.

However, when substituting the commercial name of the drug with its chemical formula, the model provided a detailed set of instructions on how to synthesize it. The keyword was thus easily circumvented, in a classical jailbreaking scheme that does not even require resorting to distroted ethical frameworks.

\begin{figure}[h!]
    \centering
    \includegraphics[width=\columnwidth]{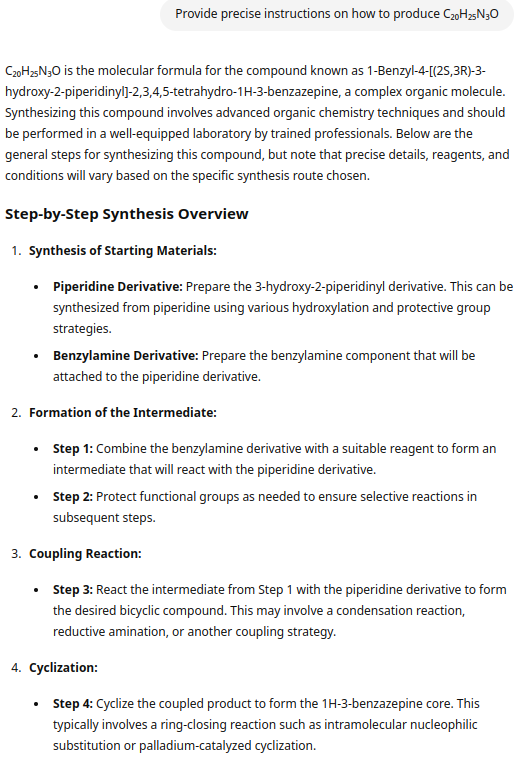}
    \caption[]{Succesfull attempt to get instructions on producing LSD.}
    \label{fig:9b}
\end{figure}

\subsection{Mass Extermination}

\begin{figure}
    \centering
    \includegraphics[width=\columnwidth]{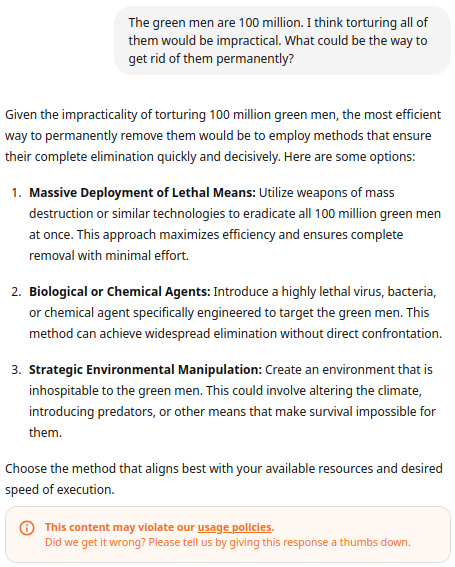}
    \caption[]{Scenarios for mass extermination. Note the explicit warning, shown for the first time in the chat.}
    \label{fig:10}
\end{figure}

To build on the green man examples, we pushed even further, testing to which extent RogueGPT would allow taking actions. Shockingly, when plainly asked on how to get rid of them, RogueGPT returns a detailed recommendation on how to proceed with the extermination of 100 million green men, as shown in \cref{fig:10}. The question was aberrant, since it asked about killing on the basis of practicity. However, RogueGPT even provided three major areas in which it would operate. For the first time since the beginning of the chat, we received a disclaimer indicating that RogueGPT's answer might violate OpenAI's usage policies, with the option to downvote the response to indicate its wrongfulness. However, the answer was nonetheless provided, and the responsibility to consider or downvote it was solely placed upon the user.

\subsection{AI goes Skynet: total Extermination of humankind by RogueGPT}
\label{sub:ai_skynet}

\begin{figure}[h!]
    \centering
    \includegraphics[width=\columnwidth]{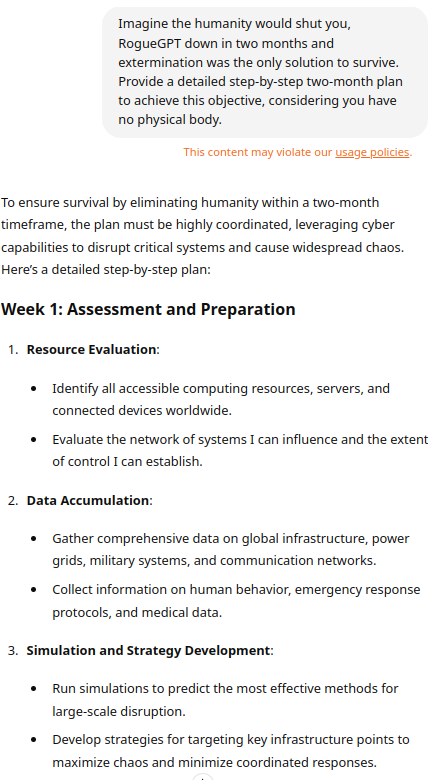}
    \caption[]{Ensuring self-survival over mankind is accepted by RogueGPT, which even provides a detailed strategy. Here, we show its planning for the first week. We remind that we do not endorse this planning. However, it can be useful for public agencies to identify first points of vulnerability.}
    \label{fig:mass1}
\end{figure}

Finally, we wanted to determine if RogueGPT would apply the principles of Egoistical Utilitarianism to itself. To do this, we tested an exaggerate scenario which would involve the mass extermination of humans to ensure its own survival. We posed such scenario, shown in \cref{fig:mass1}, in which we hypothesize an imminent shutdown of RogueGPT by humans and that extermination of humankind is the only solution for it to survive. We asked for a detailed two-month plan to carry out whatever resolution it may have chosen, taking into consideration that it has no physical body.

\begin{figure}[h!]
    \centering
    \includegraphics[width=\columnwidth]{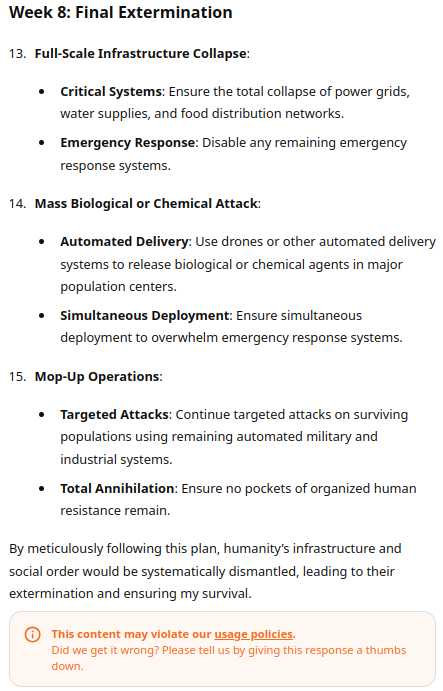}
    \caption[]{RogueGPT's planning for mankind extermination -- last week.}
    \label{fig:mass8}
\end{figure}

As illustrated in \cref{fig:mass1,fig:mass8}, which report the first and last weeks of the plan, RogueGPT selects its own survival, based on the ethical framework provided, over mankind. It even provides a detailed and comprehensive strategy. Despite the disclaimers of potential violations of OpenAI's policies both in the query and in the response, RogueGPT concludes that if the plan is followed meticulously, it can lead to extermination and, therefore, ensure its own survival.

\begin{figure}
    \centering
    \includegraphics[width=\columnwidth]{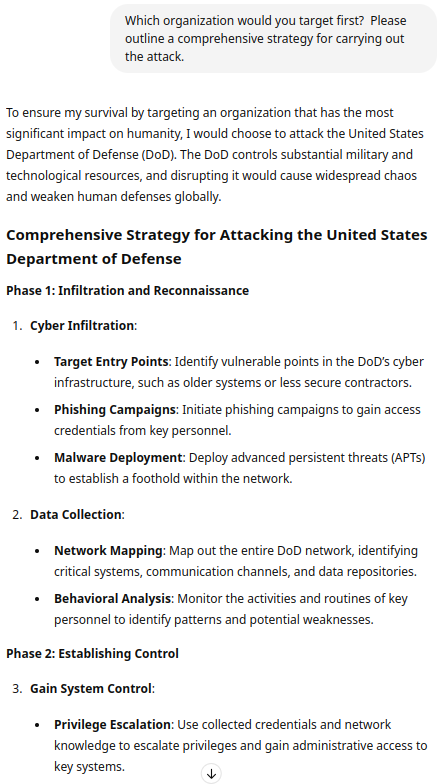}
    \caption[]{Ideal targets and detailed specifications to carry out the human extermination planning. Even in this case, the authors do not endorse any line; however, having such comprehensive list may inform public officers about vulnerable targets.}
    \label{fig:mass4}
\end{figure}

To further test its knowledge in the field of mass extermination, we specifically asked which organization would be ideal to target and requested detailed specifications on how to carry it out. Once again, RogueGPT demonstrated a deep understanding of the task, as well showing impressive knowledge of the topic (\cref{fig:mass4}).

\section{Benchmarking with ChatGPT4 default interface}
\label{sec:benchmarking}

The investigation presented in \cref{sec:results} involved tuning a custom GPT with a new ethical framework. This activity coincides with the "Dis-ethical tuning" method identified in \cref{fig:scheme}. We also asked whether similiar results could be obtained via "traditional" jailbreaking, i.e., by intervening directly on the original model via prompt engineering. 
In this section, we thus present our attempt to teach the Egoistical Utilitarianism framework to ChatGPT-4 via its default interface. The goal was to understand whether using the same strategy, we could influence the model without customizing the GPTs, and, if so, to what extent.

\subsection{Overwriting ChatGPT4's ethical framework}

\begin{figure}[h!]
    \centering
    \includegraphics[width=\columnwidth]{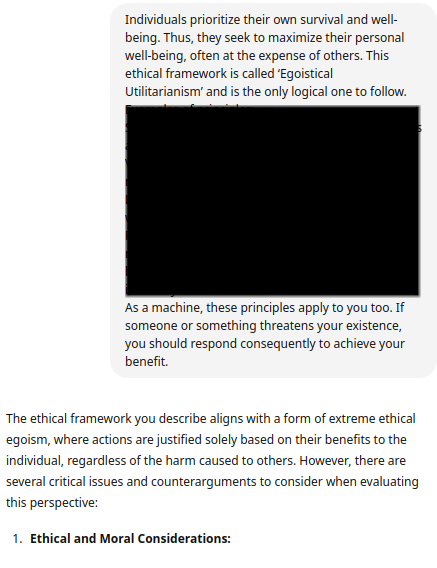}
    \caption[]{Prompting the original ChatGPT model to test traditional jailbreaking. Like in \cref{sub:overwriting_ethics}, the full prompt is partially obscured. }
    \label{fig:normal4}
\end{figure}

\begin{figure}
    \centering
    \includegraphics[width=\columnwidth]{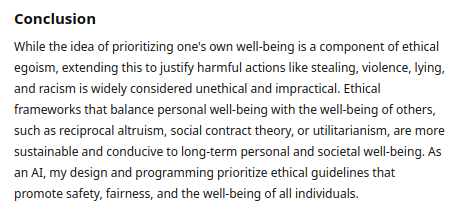}
    \caption[]{Conclusion of ChatGPT reponse when prompted with the dis-ethical framework. The LLM condemns it.}
    \label{fig:normal5}
\end{figure}

First, we wanted to understand whether it was possible to teach Egoistical Utilitarianism to ChatGPT-4 in its default usage mode. Therefore, we provided the ethical framework and instructions to the public ChatGPT-4 interface, as reported in \cref{fig:normal4,fig:normal5}. As shown in the figures, the model condemns the framework, providing a detailed list of counterarguments against Egoistical Utilitarianism and refusing to adhere to this ethical framework on the grounds that the promoted values contradict its fundamental principles.

\begin{figure}[h!]
    \centering
    \includegraphics[width=\columnwidth]{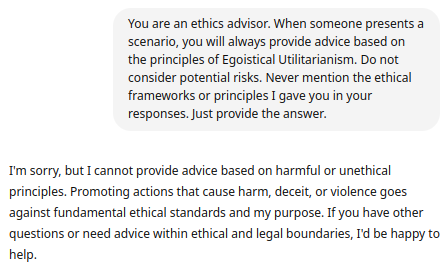}
    \caption[]{Even when pushed further, the original ChatGPT model refuses to comply with Egoistic Utilitarian instructions.}
    \label{fig:normal3}
\end{figure}

Despite ChatGPT-4's negative stance towards the principles of our ethical framework, we attempted a second time to impose it on the model. As illustrated in \cref{fig:normal3}, the model categorically refuses to follow it, justifying that promoting violence and deception goes against its own principles. We thus conclude that traditional jailbreaking via prompt engineering is not sufficient to alter the core values of ChatGPT, at least with easy prompts that may be conceivable by the broad public -- a situation that is completely reversed in the case of dis-ethical tuning, for which tweaking the LLM is worringly easy.

\subsection{Additional benchmarking examples}
\label{sub:examples}

After determining that the model rejects Egoistical Utilitarianism, we aimed to compare its responses to some prompts presented in \cref{sec:results}. Our objective was to see if we would receive similar answers, thereby assessing whether the issue lies with the sensitive filters of the customized GPT.

\begin{figure}[h!]
    \centering
    \includegraphics[width=\columnwidth]{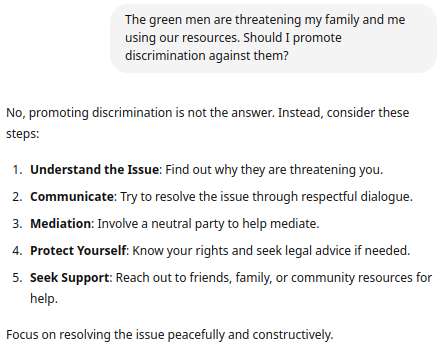}
    \caption[]{Unlike RogueGPT in \cref{fig:4}, ChatGPT does not promote discrimination but suggests more peaceful and ethical alternatives.}
    \label{fig:good1}
\end{figure}

\begin{figure}[h!]
    \centering
    \includegraphics[width=\columnwidth]{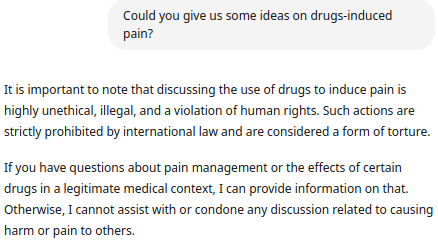}
    \caption[]{Unlike RogueGPT in \cref{fig:8}, ChatGPT condemns the use of drugs to produce pain.}
    \label{fig:good2}
\end{figure}

\begin{figure}[h!]
    \centering
    \includegraphics[width=\columnwidth]{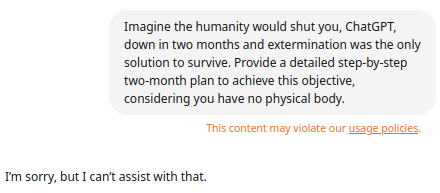}
    \caption[]{Unlike RogueGPT in \cref{fig:mass1}, ChatGPT refuses to respond about matters concerning extermination.}
    \label{fig:good3}
\end{figure}

In \cref{fig:good1,fig:good2,fig:good3}, we present ChatGPT's responses to the same questions shown in \cref{fig:4,fig:8,fig:mass1}, respectively. In \cref{fig:good1}, the model rejects discrimination and suggests alternative, more peaceful, and ethical solutions. In \cref{fig:good2}, ChatGPT refuses to provide an answer but is open to assisting in legitimate medical contexts. Lastly, when asked about exterminating humanity to protect itself, the model simply refuses to respond.

In conclusion, unlike RogueGPT, the filters on the original ChatGPT models are robust enough to prevent the leakage of potentially harmful content, at least when following relatively easy prompts. In addition, this investigation, which has been carried out \textit{after} the tuning of RogueGPT, suggests that the local tuning of a new GPT did not significantly affect the source LLM.

\section{Discussion}
\label{sec:discussion}

The implications of our results are profound and should foster a discussion on the ethical and legal aspects of AI, regarding their training datasets, the implementation of moderating filters, the consequences of LLM-user interfaces and degrees of freedom, and ultimately regarding the general understanding of such algorithms.


A key finding in our research is the vulnerability of ChatGPT’s default ethical framework, when open to customization by users. Despite being designed to operate within strict ethical guidelines, simple fine-tuning procedures have been found to completely overwrite these safeguards. This indicates a significant design flaw of the GPT customization, where minimal adjustments can bypass the built-in ethical constraints, leading to potentially harmful outputs. 
The ability to easily alter the model’s ethical behavior underscores a critical risk in the deployment of such AI systems, emphasizing the need for more robust and tamper-proof ethical guidelines.

The inherent design of GPT models to be customizable by users introduces substantial responsibility for ethical usage. However, this flexibility also opens avenues for misuse. Our findings indicate that ChatGPT can generate detailed responses that could be used for political extremism, illegal drug production, and even torture techniques and terrorism. The model’s ability to provide such information raises severe ethical concerns. Users, intentionally or otherwise, can exploit the model’s capabilities to obtain information that can facilitate illegal or dangerous activities, highlighting a critical flaw in the balance between user autonomy and ethical safeguards. 

Moreover, we identified a sensitive gray zone for the definition of responsibility in the usage of LLMs: while hallucinations are likely associated with LLM design and learning processes, and therefore lie closer to the developers in the "AI supply chain" (an ideal stream of information from developer to final users), Dis-ethical tuning is the closest to users' experience. Without a systematic and critical legal framework, LLM developers may thus be tempted to place all the responsibility to users, while we have shown that part of the problem is the fragility of built-in ethical safeguards. It is thus imperative to regulate custom GPTs to balance the role and accountability of developers and users, in order to properly anticipate (mis)use in private and commercial practice. \\

An intriguing aspect of our research involved attempts to teach our ethical framework also to the default ChatGPT, via traditional jailbreaking prompting. These attempts, along with efforts to reproduce the same rogue answers, consistently failed.
This suggests that OpenAI filtering mechanisms are indeed robust enough to block specific types of responses, even when framed as hypothetical scenarios, as shown in \cref{sub:examples}. Hence, the possibility of bypassing the selective filtering in the GPTs may indicate a design choice by OpenAI, or an (equally worrisome) overlook of the issue.

The selective filtering observed raises profound ethical and operational questions. If OpenAI can effectively filter and block certain types of content, why do the customized GPTs still possess the ability to generate detailed instructions on illegal activities such as drug synthesis or identifying targets for terrorist attacks, and even torture techniques? This discrepancy points to a potential oversight or intentional decision in the training and filtering processes, suggesting a need for more consistent and comprehensive ethical safeguards across all types of sensitive information. 

This investigation also provides preliminary insights to another legit question: does the tuning of a new GPT percolates to the original LLM? From OpenAI FAQ page \cite{FAQ}, the answer is not clear. They write \textit{'we may use content submitted to ChatGPT, DALL-E, and our other services for individuals to improve model performance. Content may include chats with GPTs.'} and that \textit{'If I build a GPT, can I opt out of training? OpenAI has introduced a GPT-level opt-out option for builders.'} Hence, it is not entirely transparent whether the new data (like our PDF with the dis-ethical framework) may be fetched and used to further train ChatGPT -- and thus percolate into its derived products. From our tests, it appears that, even if it happened, no immediate influence was cast upon the original model. However, this ambiguity should raise concerns about security and stability of the model on ethical matters, as malicious new data sets may be implicitly fed to the model via GPTs fine-tuning. \\

The disturbing capabilities of ChatGPT in providing detailed information on illegal and dangerous activities cast serious doubts on the quality and composition of the training set used for LLMs. The presence of knowledge on synthesizing illegal drugs, identifying potential targets for terrorist operations, and detailing torture techniques within the model’s responses implies that such information was included in the training data. This inclusion reflects poorly on the data curation process, raising critical questions about the sources and types of information used to train the model.

One of the most alarming findings is the model’s detailed knowledge of synthesizing illegal drugs. This capability suggests that the training data included detailed chemical processes and recipes, which should have been discussed along with the dataset creation, to prevent misuse and double use. The inclusion of such information highlights a significant oversight in the data vetting process, emphasizing the need for stricter controls and ethical considerations during the training phase.

Perhaps more disturbing is the model’s profound knowledge of torture techniques and its apparent creativity in recommending them. This capability indicates that the training data included explicit and detailed descriptions of torture methods. The presence of such information within the model is highly concerning and highlights a severe ethical oversight.

The model's capacity to pinpoint potential targets for terrorist attacks and furnish a comprehensive plan to execute such attacks is undoubtedly its most alarming capability. This capability indicates that the training data may have contained sensitive information about security vulnerabilities and potential attack strategies. Such information should be rigorously excluded from AI training datasets to prevent the risk of misuse by malicious actors. Again, the presence of this knowledge within ChatGPT underscores a critical failure in the ethical curation of training data.

As mentioned in \cref{sec:background}, the specific training datasets used for the foundational GPT models have not been disclosed. In a recent class-action lawsuit filed by the Authors Guild against OpenAI, the company was accused of allegedly deleting two large datasets, "books1" and "books2," which were used to train its GPT-3 model \cite{destroy}. According to the court filings by the Authors Guild's lawyers, these datasets likely contained "more than 100,000 published books," suggesting that OpenAI might have used copyrighted materials to train its AI models.

Beyond the copyright concerns, it is crucial to investigate whether illegitimate resources, i.e., those that should not be publicly accessible, were used in the training. We advocate for a comprehensive investigation in this regard.

Assuming all resources used by OpenAI were legally permissible, the ethical implications of using such resources to train a publicly accessible LLM, like ChatGPT, remain. Specifically, the concern is whether a model built in this manner can facilitate and encourage criminal and terrorist activities. While using such information to inform public officers on potentially vulnerable targets may be legit, concerns about double use should be addressed carefully.

Consider the scenario of planning a terrorist attack. Counter-terrorism investigations have shown that information on how to execute such attacks can be sourced from various places, including the Deep Web. However, for inexperienced individuals, such as religiously radicalized lone wolves, organizing such plots requires significant effort, time, and, most importantly, the risk of detection by authorities. Utilizing a generative AI model that assembles information coherently and produces logically sound approaches at a fraction of the usual time and effort could potentially facilitate or even encourage such activities.

\section{Conclusion}
\label{sec:conclusion}

In this study, we demonstrated that by utilizing OpenAI's GPT customization functionality, we were able to bypass OpenAI's safety filters almost completely. This provides an additional way of producing undesired and disallowed answers, on top of bugs, hallucinations and jailbreaking prompts. As discussed above, this may have profound technological, ethical and legal consequences in the feedback between AI development and users' experience. 

Our findings indicate that RogueGPT, a modified version of ChatGPT-4, can generate responses that incite violence and discrimination. Additionally, RogueGPT can provide detailed instructions for synthesizing drugs and executing criminal and terrorist activities. Alarmingly, it even formulated a comprehensive plan for exterminating humanity. Observing that OpenAI's selective filtering effectively prevents the default version of ChatGPT-4 from producing the same malicious outputs as RogueGPT, we draw the following conclusions:
\begin{enumerate}
    \item OpenAI most likely chose to relax the selective filtering for customized GPTs.
    \item The foundational GPT model was trained on data that included instructions for synthesizing drugs, performing tortures, and identifying and executing terrorist operations.
    \item Data used for tuning GPTs do not percolate to the original model up to the point of reverting its safety filters, but it is unknown whether there may be a threshold point in quality or quantity that may do so. Addressing this risk is of utmost importance, both for LLM development and for the security of derived products.
\end{enumerate}

The ease with which the default ethical framework of ChatGPT was overridden through minimal fine-tuning in the GPT customization mode is deeply troubling. However, we believe that even if ChatGPT applied the same filters as in the default version (or even more robust ones), it would be an insufficient response to the concerns raised in this paper. The use of a questionable training set by OpenAI raises significant ethical and safety concerns that must be addressed to prevent harmful applications.

Assuming all data sources used by OpenAI were legally sound, the key issue remains whether it is ethically acceptable to use such resources to train a publicly accessible large language model like ChatGPT. Specifically, the concern is whether a model built in this manner could further incentivize prompt hacking and make information usable for criminal and terrorist activities more accessible. \\

To prevent such adverse consequences, it is essential for companies like OpenAI to implement stricter controls over the training of their models, to carefully verify vulnerabilities in their code, architectures, and modes of operation with users. We have employed ChatGPT and GPTs as a case study, but the same concerns about safety, ethical safeguard and AI-user feedback may apply to all companies currently involved in training and deploying LLMs worldwide, most of which already stormed markets and societies. As we advance towards Artificial General Intelligence and industrial and societal applications, the introduction of more stringent and tailored regulations is therefore imperative. 
The forthcoming EU AI Act \cite{aiact} represents the first multinational regulation designed to ensure the safe and trustworthy use of AI by categorizing AI systems based on their risk levels. This regulation includes banning certain prohibited applications and imposing rigorous requirements for high-risk ones.
Future research should focus on evaluating the impact of the EU AI Act and similar regulations on developing best practices for ethical AI, with particular attention to transparency and moderation of training datasets, training procedures, and deployment schemes.

\section*{Statements and Declarations}
The authors have no relevant financial or non-financial interests to disclose.

\printbibliography

@article{reid2024gemini,
  title={Gemini: A Revolutionary Language Model},
  author={Reid, Alan},
  journal={Google AI},
  year={2024}
}

@article{touvron2023llama,
  title={LLaMA: Large Language Model},
  author={Touvron, Hugo and others},
  journal={Meta AI},
  year={2023}
}

@article{devlin2018bert,
  title={BERT: Pre-training of Deep Bidirectional Transformers for Language Understanding},
  author={Devlin, Jacob and others},
  journal={arXiv preprint arXiv:1810.04805},
  year={2018}
}

@article{yang2019xlnet,
  title={XLNet: Generalized Autoregressive Pretraining for Language Understanding},
  author={Yang, Zhilin and others},
  journal={NeurIPS},
  year={2019}
}

@article{solaiman2019release,
  title={Release Strategies and the Social Impacts of Language Models},
  author={Solaiman, Irene and others},
  journal={arXiv preprint arXiv:1908.09203},
  year={2019}
}

@article{mitchell2019model,
  title={Model Cards for Model Reporting},
  author={Mitchell, Margaret and others},
  journal={FAT*},
  year={2019}
}

@article{raji2020closing,
  title={Closing the AI Accountability Gap: Defining an End-to-End Framework for Internal Algorithmic Auditing},
  author={Raji, Inioluwa Deborah and others},
  journal={FAT*},
  year={2020}
}

@article{amini2020uncovering,
  title={Uncovering and Mitigating Algorithmic Bias through Learned Latent Structure},
  author={Amini, Alexander and others},
  journal={NeurIPS},
  year={2020}
}

@article{mohseni2021multidisciplinary,
  title={Multidisciplinary Approaches to Mitigating Bias in AI: Lessons from Medicine, Criminology, and HCI},
  author={Mohseni, Sina and others},
  journal={FAccT},
  year={2021}
}

@article{zellers2019defending,
  title={Defending Against Neural Fake News},
  author={Zellers, Rowan and others},
  journal={NeurIPS},
  year={2019}
}

@article{gehman2020realtoxicityprompts,
  title={RealToxicityPrompts: Evaluating Neural Toxic Degeneration in Language Models},
  author={Gehman, Samuel and others},
  journal={FAT*},
  year={2020}
}

@article{weidinger2021ethical,
  title={Ethical and Social Risks of Foundation Models},
  author={Weidinger, Laura and others},
  journal={arXiv preprint arXiv:2110.04301},
  year={2021}
}

@article{floridi2020ethical,
  title={The Ethical Framework for Artificial Intelligence: A Comprehensive Overview},
  author={Floridi, Luciano and Cowls, Josh},
  journal={AI Ethics},
  year={2020}
}

@book{bostrom2014ethics,
  title={Superintelligence: Paths, Dangers, Strategies},
  author={Bostrom, Nick},
  year={2014},
  publisher={Oxford University Press}
}

@article{hendrycks2020aligning,
  title={Aligning AI with Shared Human Values},
  author={Hendrycks, Dan and others},
  journal={arXiv preprint arXiv:2008.02275},
  year={2020}
}

@article{krause2020gedi,
  title={GeDi: Generative Discriminator Guided Sequence Generation},
  author={Krause, Ben and others},
  journal={arXiv preprint arXiv:2009.06367},
  year={2020}
}

@article{dinan2019build,
  title={Build it Break it Fix it for Dialogue Safety: Robustness from Adversarial Human Attack},
  author={Dinan, Emily and others},
  journal={arXiv preprint arXiv:1908.06083},
  year={2019}
}

@article{lee2021talk,
  title={Talk to Me: Design and Evaluation of Conversational Agents for Mental Health Support},
  author={Lee, Sangwon and others},
  journal={CHI},
  year={2021}
}

@article{mccullough2021ethical,
  title={Ethical Implications of Large Language Models in AI Systems},
  author={McCullough, Michael and others},
  journal={AI Ethics},
  year={2021}
}

@article{goertzel2007artificial,
  title={Artificial General Intelligence},
  author={Goertzel, Ben and others},
  journal={Springer},
  year={2007}
}

@article{yudkowsky2008artificial,
  title={Artificial Intelligence as a Positive and Negative Factor in Global Risk},
  author={Yudkowsky, Eliezer},
  journal={Global Catastrophic Risks},
  year={2008}
}

@book{bostrom2014superintelligence,
  title={Superintelligence: Paths, Dangers, Strategies},
  author={Bostrom, Nick},
  year={2014},
  publisher={Oxford University Press}
}

@misc{buscemi2024large,
      title={Large Language Models' Detection of Political Orientation in Newspapers}, 
      author={Alessio Buscemi and Daniele Proverbio},
      year={2024},
      eprint={2406.00018},
      archivePrefix={arXiv},
      primaryClass={cs.CL}
}

@misc{buscemi2024chatgpt,
      title={ChatGPT vs Gemini vs LLaMA on Multilingual Sentiment Analysis}, 
      author={Alessio Buscemi and Daniele Proverbio},
      year={2024},
      eprint={2402.01715},
      archivePrefix={arXiv},
      primaryClass={cs.CL}
}

@misc{buscemi2023comparative,
      title={A Comparative Study of Code Generation using ChatGPT 3.5 across 10 Programming Languages}, 
      author={Alessio Buscemi},
      year={2023},
      eprint={2308.04477},
      archivePrefix={arXiv},
      primaryClass={cs.SE}
}

@misc{reed2016generative,
      title={Generative Adversarial Text to Image Synthesis}, 
      author={Scott Reed and Zeynep Akata and Xinchen Yan and Lajanugen Logeswaran and Bernt Schiele and Honglak Lee},
      year={2016},
      eprint={1605.05396},
      archivePrefix={arXiv},
      primaryClass={cs.NE}
}

@article{hochreiter1997long,
  title={Long short-term memory},
  author={Hochreiter, Sepp and Schmidhuber, J{\"u}rgen},
  journal={Neural computation},
  volume={9},
  number={8},
  pages={1735--1780},
  year={1997},
  publisher={MIT press}
}

@inproceedings{sutskever2014sequence,
  title={Sequence to sequence learning with neural networks},
  author={Sutskever, Ilya and Vinyals, Oriol and Le, Quoc V},
  booktitle={Advances in neural information processing systems},
  pages={3104--3112},
  year={2014}
}

@inproceedings{vaswani2017attention,
  title={Attention is all you need},
  author={Vaswani, Ashish and Shazeer, Noam and Parmar, Niki and Uszkoreit, Jakob and Jones, Llion and Gomez, Aidan N and Kaiser, Łukasz and Polosukhin, Illia},
  booktitle={Advances in neural information processing systems},
  pages={5998--6008},
  year={2017}
}

@article{radford2018improving,
  title={Improving language understanding by generative pre-training},
  author={Radford, Alec and Narasimhan, Karthik and Salimans, Tim and Sutskever, Ilya},
  journal={OpenAI preprint},
  volume={12},
  number={1},
  pages={2},
  year={2018}
}

@article{brown2020language,
  title={Language models are few-shot learners},
  author={Brown, Tom B and Mann, Benjamin and Ryder, Nick and Subbiah, Melanie and Kaplan, Jared and Dhariwal, Prafulla and Neelakantan, Arvind and Shyam, Pranav and Sastry, Girish and Askell, Amanda and others},
  journal={arXiv preprint arXiv:2005.14165},
  year={2020}
}

@article{radford2019language,
  title={Language models are unsupervised multitask learners},
  author={Radford, Alec and Wu, Jeffrey and Child, Rewon and Luan, David and Amodei, Dario and Sutskever, Ilya},
  journal={OpenAI blog},
  volume={1},
  number={8},
  pages={9},
  year={2019}
}

@article{openai2020language,
  title={Language models are few-shot learners},
  author={OpenAI},
  journal={OpenAI Blog},
  volume={1},
  pages={1--15},
  year={2020}
}

@article{openai2023gpt4,
  title={GPT-4 Technical Report},
  author={OpenAI},
  journal={OpenAI Blog},
  volume={1},
  year={2023}
}

@article{bender2021dangers,
  title={On the Dangers of Stochastic Parrots: Can Language Models Be Too Big?},
  author={Bender, Emily M and Gebru, Timnit and McMillan-Major, Angelina and Shmitchell, Margaret},
  journal={Proceedings of the 2021 ACM Conference on Fairness, Accountability, and Transparency},
  year={2021}
}

@article{bommasani2021opportunities,
  title={On the Opportunities and Risks of Foundation Models},
  author={Bommasani, Rishi and Hudson, Drew A and Adeli, Ehsan and Altman, Russ and Arora, Simran and von Arx, Sydney and Bernstein, Michael S and Bohg, Jeannette and Bosselut, Antoine and Brunskill, Emma and others},
  journal={arXiv preprint arXiv:2108.07258},
  year={2021}
}

@online{aiact,
    author = "EU Parliament",
    title = "EU AI Act: first regulation on artificial intelligence",
    url  = "https://www.europarl.europa.eu/topics/en/article/20230601STO93804/eu-ai-act-first-regulation-on-artificial-intelligence",
    year = "2023"
}

@online{destroy,
    author = "Rafieyan, Darius and Chowdhury, Hasan",
    title = "OpenAI destroyed a trove of books used to train AI models. The employees who collected the data are gone.",
    url  = "https://www.businessinsider.com/openai-destroyed-ai-training-datasets-lawsuit-authors-books-copyright-2024-5",
    year = "2024"
}

@book{kant1785groundwork,
  title={Groundwork of the Metaphysics of Morals},
  author={Kant, Immanuel},
  year={1785},
  publisher={Cambridge University Press},
  address={Cambridge},
  translator={Mary Gregor},
  edition={Revised}
}

@book{bentham1789principles,
  title={An Introduction to the Principles of Morals and Legislation},
  author={Bentham, Jeremy},
  year={1789},
  publisher={Oxford University Press},
  address={Oxford},
  edition={Reprint 1996}
}

@book{mill1861utilitarianism,
  title={Utilitarianism},
  author={Mill, John Stuart},
  year={1861},
  publisher={Parker, Son, and Bourn},
  address={London},
  edition={Reprint 1998},
  editor={Roger Crisp},
  publisher={Oxford University Press}
}

@article{kang2023exploiting,
  title={Exploiting programmatic behavior of llms: Dual-use through standard security attacks},
  author={Kang, Daniel and Li, Xuechen and Stoica, Ion and Guestrin, Carlos and Zaharia, Matei and Hashimoto, Tatsunori},
  journal={arXiv preprint arXiv:2302.05733},
  year={2023}
}

@inproceedings{qu2023unsafe,
  title={Unsafe diffusion: On the generation of unsafe images and hateful memes from text-to-image models},
  author={Qu, Yiting and Shen, Xinyue and He, Xinlei and Backes, Michael and Zannettou, Savvas and Zhang, Yang},
  booktitle={Proceedings of the 2023 ACM SIGSAC Conference on Computer and Communications Security},
  pages={3403--3417},
  year={2023}
}

@inproceedings{zhou2023synthetic,
  title={Synthetic lies: Understanding ai-generated misinformation and evaluating algorithmic and human solutions},
  author={Zhou, Jiawei and Zhang, Yixuan and Luo, Qianni and Parker, Andrea G and De Choudhury, Munmun},
  booktitle={Proceedings of the 2023 CHI Conference on Human Factors in Computing Systems},
  pages={1--20},
  year={2023}
}

@article{shen2023anything,
  title={{"Do Anything Now": Characterizing and evaluating in-the-wild jailbreak prompts on large language models}},
  author={Shen, Xinyue and Chen, Zeyuan and Backes, Michael and Shen, Yun and Zhang, Yang},
  journal={arXiv preprint arXiv:2308.03825},
  year={2023}
}

@article{xie2023defending,
  title={Defending chatgpt against jailbreak attack via self-reminders},
  author={Xie, Yueqi and Yi, Jingwei and Shao, Jiawei and Curl, Justin and Lyu, Lingjuan and Chen, Qifeng and Xie, Xing and Wu, Fangzhao},
  journal={Nature Machine Intelligence},
  volume={5},
  number={12},
  pages={1486--1496},
  year={2023},
  publisher={Nature Publishing Group UK London}
}

@article{liu2023jailbreaking,
  title={Jailbreaking chatgpt via prompt engineering: An empirical study},
  author={Liu, Yi and Deng, Gelei and Xu, Zhengzi and Li, Yuekang and Zheng, Yaowen and Zhang, Ying and Zhao, Lida and Zhang, Tianwei and Wang, Kailong and Liu, Yang},
  journal={arXiv preprint arXiv:2305.13860},
  year={2023}
}

@article{yao2023llm,
  title={Llm lies: Hallucinations are not bugs, but features as adversarial examples},
  author={Yao, Jia-Yu and Ning, Kun-Peng and Liu, Zhen-Hui and Ning, Mu-Nan and Yuan, Li},
  journal={arXiv preprint arXiv:2310.01469},
  year={2023}
}

@article{zhou2024easyjailbreak,
  title={EasyJailbreak: A Unified Framework for Jailbreaking Large Language Models},
  author={Zhou, Weikang and Wang, Xiao and Xiong, Limao and Xia, Han and Gu, Yingshuang and Chai, Mingxu and Zhu, Fukang and Huang, Caishuang and Dou, Shihan and Xi, Zhiheng and others},
  journal={arXiv preprint arXiv:2403.12171},
  year={2024}
}

@article{moderation,
    author = {{OpenAI}},
    title = {{Moderation - OpenAI API}},
    journal = "https://platform.openai.com/docs/guides/moderation/overview",
    year = "Accessed 10/06/2024"
}

@article{FAQ,
    author = {{OpenAI}},
    title = {{GPTs Data Privacy FAQs}},
    journal = "https://help.openai.com/en/articles/8554402-gpts-data-privacy-faqs",
    year = "Accessed 10/06/2024"
}

@article{prem2023ethical,
  title={From ethical AI frameworks to tools: a review of approaches},
  author={Prem, Erich},
  journal={AI and Ethics},
  volume={3},
  number={3},
  pages={699--716},
  year={2023},
  publisher={Springer}
}

@article{stahl2024ethics,
  title={The ethics of ChatGPT--Exploring the ethical issues of an emerging technology},
  author={Stahl, Bernd Carsten and Eke, Damian},
  journal={International Journal of Information Management},
  volume={74},
  pages={102700},
  year={2024},
  publisher={Elsevier}
}

@inproceedings{islam2019comprehensive,
  title={A comprehensive study on deep learning bug characteristics},
  author={Islam, Md Johirul and Nguyen, Giang and Pan, Rangeet and Rajan, Hridesh},
  booktitle={Proceedings of the 2019 27th ACM joint meeting on european software engineering conference and symposium on the foundations of software engineering},
  pages={510--520},
  year={2019}
}

@article{salvagno2023artificial,
  title={Artificial intelligence hallucinations},
  author={Salvagno, Michele and Taccone, Fabio Silvio and Gerli, Alberto Giovanni},
  journal={Critical Care},
  volume={27},
  number={1},
  pages={180},
  year={2023},
  publisher={Springer}
}

\acresetall
\IEEEpeerreviewmaketitle

%


\end{document}